\documentclass[twocolumn,showpacs,preprintnumbers,amsmath,amssymb]{revtex4}
\usepackage{graphicx}% Include figure files
\usepackage{dcolumn}% Align table columns on decimal point
\usepackage{bm}% bold math
\usepackage{makecell} %for boxes inside the table
\usepackage[english]{babel}
\usepackage{verbatim}
\usepackage{color}
\begin{document}
%\begin{comment}
\preprint{}

\title{Electrical properties of in-plane-implanted graphite nanoribbons}% Force line breaks with \\
%Effects of $\text{Ga}^{+}$ implantation perpendicular to c-axis in thin graphite ribbons

\author{B. C. Camargo}
 \email{b.c_camargo@yahoo.com.br}
\affiliation{%
Institute of Physics, Polish Academy of Sciences, Aleja Lotnikow 32/46, PL-02-668 Warsaw, Poland.}%
 \affiliation{Division of Superconductivity and Magnetism, Felix Bloch Institute for Solid State Physics,
Universitat Leipzig, Linn\'estrasse 5, D-04103 Leipzig, Germany }%Lines break automatically or can be forced with \\
%\author{Y. Kopelevich}%

\author{R. F. de Jesus}
\affiliation{%
Instituto de F\'isica, Universidade Federal do Rio Grande do Sul, 91501-970, Porto Alegre, RS, Brazil}%

\author{B. V. Semenenko}
\author{C. E. Precker}
%\author{P. Esquinazi}
\affiliation{Division of Superconductivity and Magnetism, Felix Bloch Institute for Solid State Physics,
Universit\"at Leipzig, Linn\'estrasse 5, D-04103 Leipzig, Germany}% with \\

\date{\today}% It is always \today, today,

\begin{abstract}
We studied the effect of low energy (30 keV) ionic implantation of $\text{Ga}^{+}$ in the direction parallel to the graphene planes (perpendicular to c-axis) in oriented graphite ribbons with widths around 500 nm. Our experiments have reproducibly shown a reduction of electrical resistance upon implantation consistent with the occurrence of ionic channeling in our devices. Our results allow for new approaches in the modulation of the charge carrier concentration in mesoscopic graphite 
\end{abstract}

\maketitle
%\end{comment}
\section{\label{intro} Introduction}
Graphite is a material composed by loosely-stacked graphene planes, bound together by Van-der-Waals forces. Due to its similarities with graphene, graphite attracts broad technological interest. Most notably, few-layer graphite has been the focus of increasingly larger research efforts in the past few years, partially due to its low electrical resistivity, high charge carrier mobility, as well as the possibilities of gap modulation and ballistic transport \cite{esquin_JAP2008_ballistic, Lui_Nature2011_tunable_gap_multigraphene, castro_2007_gap_bilayer_graphene, PRB_2008_garcia_balistic}.

One of the greatest challenges posed by mesoscopic graphite structures is the difficulty to control its charge carrier density. Due to its high native charge carrier concentration (ranging from $10^{18}$ to $10^{21}$ $\text{cm}^{-3}$ \cite{PRB_2008_garcia_balistic}), electrostatic doping is usually ineffective in samples with thickness above few nanometers because of charge screening. In addition, conventional ionic implantation is often problematic due to the disorder introduced in the graphite structure by the highly energetic ions \cite{barzola_nanotech2010_Ga_irradiation, arndt_PRB2009_implantation_R_reduction, miyazaki_APExp2008_screeniong_graphite}.

In this letter, we address these issues by attempting ionic implantation parallel to the planes (perpendicular to c-axis) in narrow HOPG (highly oriented pyrolytic graphite) ribbons. Our results show a consistent resistivity reduction with the amount of implanted ions, suggesting implantation in this direction as a viable way to modulate the charge carrier density in the material.

\section{Results and discussion}

\begin{figure}[h]
\includegraphics[width=7cm]{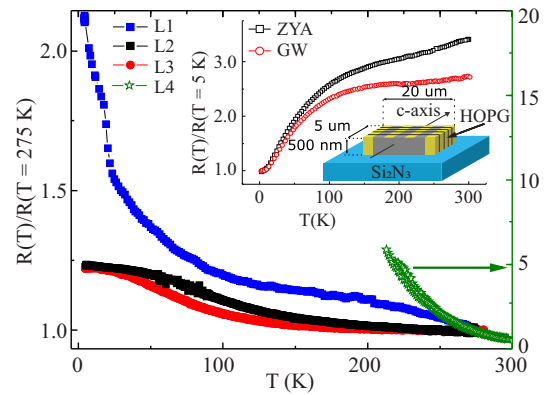}
\caption{(color online) Normalized R(T) measurements for four different ribbons. The left axis corresponds to samples L1, L2 and L3, while the right axis (green) corresponds to sample L4 (open symbols).  The ice-temperature resistance R(T=$275$ K) of the samples were $2.3$ k$\Omega$ for sample L1, $1.5$ k$\Omega$ for L2, $0.3$ k$\Omega$ for L3 and $15.0$ k$\Omega$ for L4. The inset shows the curves for the bulk HOPG used in this work. The cartoon in the inset represents the geometry of samples L1-L4, as well as the typical dimensions of the devices. In it, lighter (yellow) patches correspond to the electrical contacts in the sample.  The arrow points the sample c-axis direction.}
\label{fig_RxT}
\end{figure}

The samples studied here were HOPG ribbons extracted from two different bulk crystals: advanced ceramics ZYA HOPG (FWHM $0.5^o$) and Great Wall Inc. GW (FWHM $0.39^o$) \cite{GW_inc, ADV_ceramics}. The bulk crystals had typical dimensions of $2$ mm x $3$ mm (in-plane) x $0.5$ mm (c-axis) and room-temperature resistivities of $20$ $\mu \Omega$.cm (ZYA) and $5$ $\mu \Omega$.cm (GW). They presented a metallic-like behaviour (dR/dT$>0$) with saturation at high T, typical of well-graphitized bulk HOPG \cite{book_kelly}. This is shown in the inset of Fig. \ref{fig_RxT}. 

The ribbons were prepared as detailed in refs. \cite{barzola-intech} and \cite{Ballestar_NJP2013_SC_lamella}. In short, they were etched from a freshly-cleaved HOPG surface with the $30$ kV $\text{Ga}^+$ ion beam from a FEI dual-beam electron microscope \cite{FEI_USA}. During the milling procedure, progressively smaller ion beam currents were used for etchings done near the ribbons edges. This was performed in order to polish the sample’s surfaces and limit the Ga diffusion in the material. SRIM simulations predict that, under these conditions, the lateral penetration of $\text{Ga}^+$ in graphite should remain under $20$ nm \cite{Ballestar_NJP2013_SC_lamella, SRIM_simm}, resulting in ribbons surrounded by, at most, a 20 nm layer of amorphous carbon. To negate milling damage to the top of the ribbon, the sample was covered in-situ with an $800$ nm-thick layer of insulating PdC, obtained by electron-beam induced deposition (EBID)  \cite{Ballestar_NJP2013_SC_lamella}. The resulting samples had typical dimensions $20$ $\mu$m x $500$ nm (in-plane) x $5$ $\mu$m (along the c-axis - see the cartoon in the inset of Fig. \ref{fig_RxT}). With a micro-manipulator, the ribbons were transferred to a Si substrate coated with a $300$ nm-thick layer of insulating $\text{Si}_2\text{N}_3$. Due their narrow in-plane width, their c-axis was oriented parallel to the substrate surface. Subsequently, the samples were soldered to the substrate with four EBID PdC staircase structures, in order to allow for an effective electrical contacting. 

For the electrical addressing, the samples were submitted to a standard electronic lithography processes. In it, the devices were covered with a $1$ $\mu$m thick layer of PMMA resist, which was allowed to cure for $30$ min at $180^o$C in a furnace. The electron-litography writting was carried out with a dose of $120$ $\mu \text{C/cm}^2$ at an acceleration potential of 15 kV. After the developing process, the sample was exposed to a soft oxigen plasma (15 W) for $120$ s to remove residues from the sample surface. This was followed by the sputtering deposition of a 10 nm-thick adhesion layer of Pd, followed by a 90 nm layer of Au. Pd was chosen as the adhesion layer due its low contact resistance and wide use in multigraphene samples (see, e.g., ref. \cite{Pd_graf}), as well as its good adherence to our $\text{Si}_2\text{N}_3$ substrate. The liftoff of the resist was made with acetone. The resulting contacts short-circuited the sample along the c-axis, in an attempt to attain a homogeneous electrical current flow across the sample cross section (see the inset of Fig. \ref{fig_RxT}

In total, four samples were experimented. They are labeled L1, L2, L3 and L4. R(T) results are shown in Fig. \ref{fig_RxT}. Samples L1, L2 and L3 were measured in the interval $4$ K $\leq$ T $\leq$ $275$ K, while sample L4 was measured for $200$ K $\leq$ T $\leq$ $300$ K due to instrumentation limitations. All ribbons showed an insulating-like R(T) behavior (dR/dT$<0$), in opposition to the metallic-like dependency observed in bulk HOPG (see the inset in Fig. \ref{fig_RxT}). The ribbons’ room-temperature (RT) resistivities were $6\times 10^4$ $\mu \Omega$.cm for sample L1, $2.1\times 10^4$ $\mu \Omega$.cm for sample L2, $0.7\times 10^4$ $\mu \Omega$.cm for sample L3 and $20\times 10^4$ $\mu \Omega$.cm for sample L4. These values are three to four orders of magnitude higher than those of bulk graphite ($5$ $\mu \Omega$.cm for GW and $20$ $\mu \Omega$.cm for ZYA). 

It would be natural to attribute the ribbons' insulating-like behavior to an eventual disorder caused by $\text{Ga}^+$ ions during the milling process. However, SRIM simulations predict a diffusion of Ga in a region below $20$ nm surrounding the ribbon, which is partially removed during the plasma etching in the lithography process. These suggest that disorder on the ribbons' surface cannot be held accountable for the high resistivity of the samples. 

Instead, the insulating-like behavior and the enhanced resistivity of our devices can be understood in the context of finite-size effects. Mean-free paths of carriers in HOPG can reach values above tenths of microns, being in the same order of magnitude of the lateral width of our ribbons \cite{PRB_2008_garcia_balistic}. Experiments in constricted HOPG demonstrate that the reduction of the lateral size of crystals below $1$ $\mu$m can result in an increase of the sample resistivity up to three orders of magnitude \cite{barzola-intech}. Such increase is non-linear with the constriction size and usually depends on the device studied \cite{barzola-intech}. This aspect of our devices will be discussed elsewhere \cite{precker_paper}. In addition, the typical RT resistivities of the ribbons shown here agree with those observed in previous studies for similar samples. For example, in refs. \cite{barzola-intech, Ballestar_NJP2013_SC_lamella, ballestar_SC-sci-tech_2014}, RT resistivities of HOPG ribbons vary between $10^4$ $\mu \Omega$.cm and $10^6$ $\mu \Omega$.cm. Such devices have shown the same insulating-like R(T) behavior found here when excited with electrical currents above few nano-Amperes.

After initial characterizations, our samples were covered with a $1$ $\mu$m-thick layer of PMMA and a narrow window was patterned between the central sample electrodes with electronic lithography. The process was followed by developing and a soft plasma etching to remove PMMA residues and reduce the amorphous carbon layer coating our device. The samples were subsequently implanted with $30$ kV $\text{Ga}^+$ ions, which were available in the same FEI dual beam electron microscope employed during the etching process. Due the sample geometry, the implantation was performed with the ion beam parallel to graphene planes (perpendicular to c-axis).

\begin{figure}[h]
\includegraphics[width=7cm]{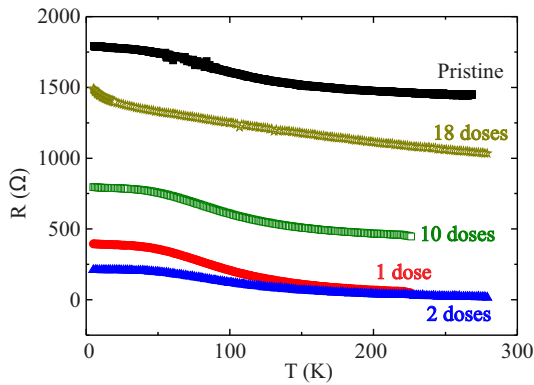}
\caption{(color online) R(T) measurements for sample L2 after consecutive ionic implantations. Note the resistance decrease, and subsequent increase, with the amount of implanted Ga.}
\label{fig_RxT_impl}
\end{figure}

\begin{figure}[h]
\includegraphics[width=7cm]{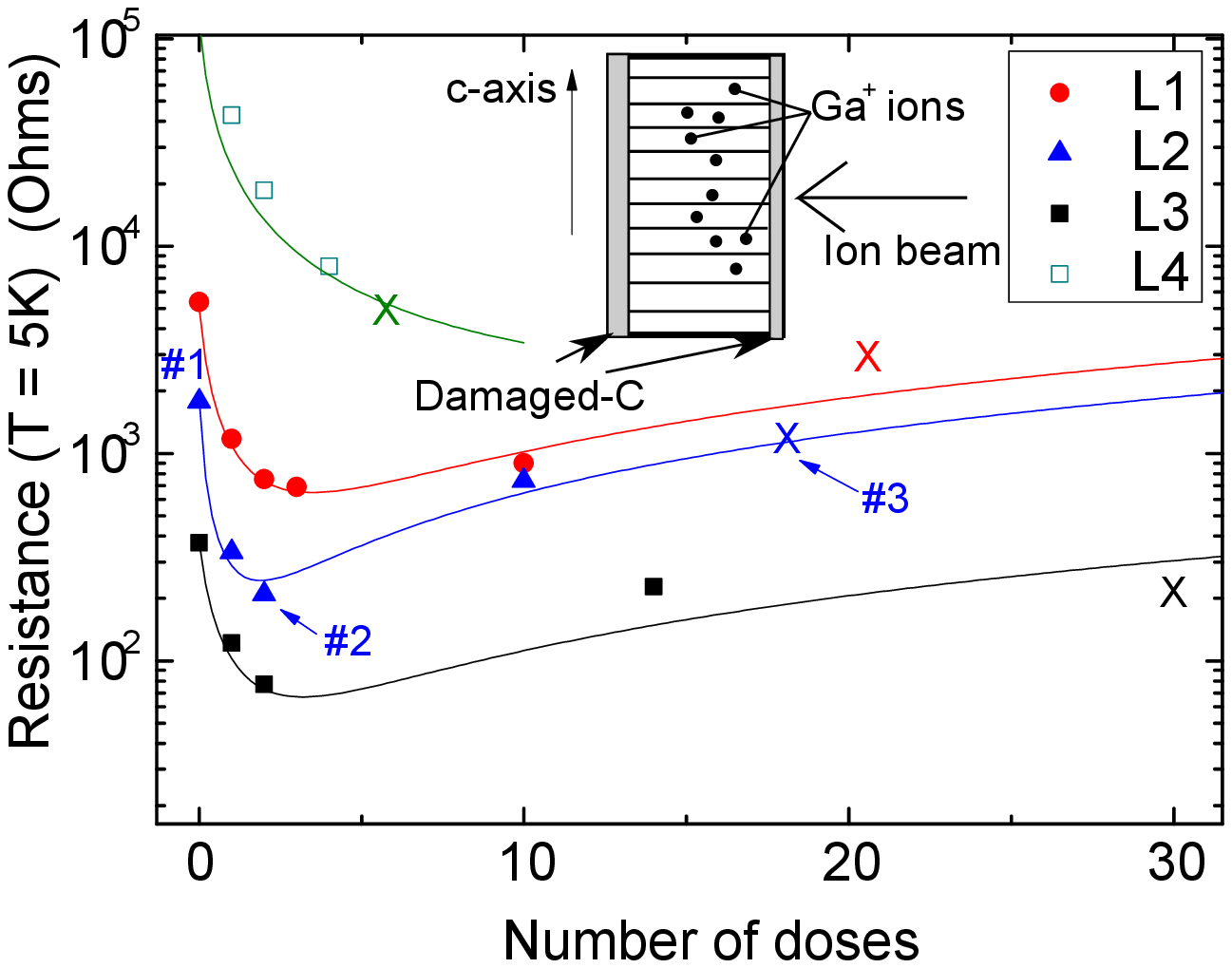}
\includegraphics[width=7cm]{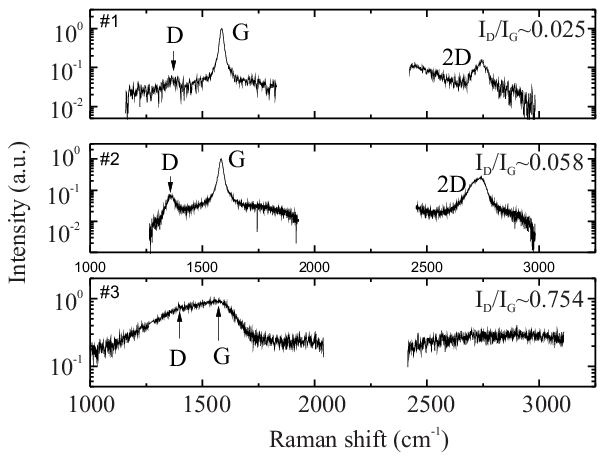}
\caption{(color online) Top Pannel: Low temperature resistance as a function of implanted Ga ions in our samples. Each dose corresponds to $2.6 \times 10^{14} \text{ ions/cm}^2$. The ``X'' corresponds to the last measured point before the samples were destroyed. The solid lines represent equation \ref{eq_impl} with the parameters $\text{R}_0 = 4.9 \times 10^3$  $\Omega$,  $\gamma =4.6 \times 10^{-3}$ and $ \beta = 90$ $\Omega \text{.cm}^2$ for sample L1, $\text{R}_0 = 1.8 \times 10^3$  $\Omega$,  $\gamma = 5.7 \times 10^{-3}$ and $ \beta = 62$ $\Omega \text{.cm}^2$ for sample L2, $\text{R}_0 = 0.38 \times 10^3$  $\Omega$,  $\gamma = 3.5 \times 10^{-3}$ and $ \beta = 20$ $\Omega \text{.cm}^2$ for sample L3 and $\text{R}_0 = 120 \times 10^3$  $\Omega$,  $\gamma = 5.4 \times 10^{-3}$ and $ \beta = 50$ $\Omega \text{.cm}^2$ for sample L4. The cartoon represents a ribbon cross-section showing how Ga was implanted in the samples. In it, The large arrow shows the ion beam orientation relative to the graphene planes in the ribbon (black transverse lines). The gray areas represent the thin amorphous C layer covering the device and the black dots the location of channeled Ga. Bottom panel: Raman measurements of sample L2 at points \#1, \#2 and \#3. Measurements were performed with the laser inciding perpendicular to the sample c-axis, with the electric field polarized along the graphene planes of the samples (similar to the Ga implantation depicted in the top panel).}
\label{fig_impl_diagram}
\end{figure}

Results for device L2 are shown in Fig. \ref{fig_RxT_impl}. In it, each implanted dose corresponds to a fluence of $2.6 \times 10^{14} \text{ ions/cm}^2$. Measurements were done ex-situ with an AC resistance bridge operating at f = $13$ Hz and excitation currents $1$ nA $\leq$ I $\leq$ $10$ nA (each curve was measured at constant excitation). All samples showed the same qualitative behavior and remained ohmic in the entire temperature range studied.  Upon Ga implantation, all devices initially became less resistive (up to one order of magnitude). Higher implanted concentrations (above $10^{15} \text{ ions/cm}^2$) caused a resitivity increase, which eventually led to the destruction of the devices. A diagram showing this effect for all our samples is presented in Fig. \ref{fig_impl_diagram} and will be discussed further ahead.

%Results have consistently shown an initial reduction of sample resistivity (up to one order of magnitude) upon increasing the amount of implanted Ga. However, higher implanted concentrations (above $10^{15} \text{ ions/cm}^2$) caused an increase of sample resistivity. Subsequent implantations caused the sample resistance to increase even further, and eventually lead to their destruction.
 
Our results seem somewhat counter-intuitive, as Ga implantation in HOPG is reported to dramatically increase the sample resistance, rather than causing the pronounced decrease observed. For example, experiments performed on thin HOPG films with the ion beam parallel to the sample c-axis have shown that implantations with doses as small as $5\times 10^{11} \text{ ions/cm}^2$ result in the increase of sample resistance above an order of magnitude \cite{barzola_nanotech2010_Ga_irradiation}. However, this behavior is not universal and strongly depends on the type of ions implanted in the material. Implantation of HOPG with $\text{H}^+$ (parallel to sample c-axis), for example, has shown a weak ($1$ \%) sample resistance reduction for an implanted dose of $10^{13} \text{ ions/cm}^2$, followed by a monotonic resistance increase at higher fluences \cite{arndt_PRB2009_implantation_R_reduction}. %Our results differ from those of reference \cite{arndt_PRB2009_implantation_R_reduction} by the type and direction of implanted ions, as well as the drastic reduction of sample resistance. In addition, in the present work, the reduction is up to one order of magnitude, in opposition to $1$\% of ref. \cite{arndt_PRB2009_implantation_R_reduction}.

The resistance increase observed in the literature is usually linked to the damage caused to in-plane chemical bondings in graphite \cite{barzola_nanotech2010_Ga_irradiation, Elman_PRB_dmg_ion_HOPG}. In our experiments, however, the ionic beam was oriented parallel to the sample planes (perpendicular to the c-axis). In this geometry, the implanted ions see atomic layers of C separated by distances about $3.35$ \AA – corresponding to the interplane distance in HOPG. These regions can allow ions to penetrate inside the material with minimal damage, in an effect known as ionic channeling.

It is well-known that both axial and planar channeling can occur in higher quality HOPG \cite{book_Dresselhaus, Elman_JAP1984_channeling_graf, SCHROYEN_1986_axial_channeling_graf, Setti_PRL1985_impl_HOPG}. Experiments measuring the backscattering of He ions in the last decade have shown that small misalignments between sample and ion beam do not considerably affect ionic channeling in the material. This applies to angles in the order of sample mosaicity, and happens due to the presence of rotational faults (twist disorder) and mosaic spread in bulk graphite \cite{Elman_PRB_dmg_ion_HOPG, book_Dresselhaus, SCHROYEN_1986_axial_channeling_graf, Meng_carbon2013_p_dop_HOPG}. In our experiments, the angle between the sample and ion beam was controlled by a step motor with a precision of $0.1^o$  - which assured that misalignment angles remained below the sample mosaicity.  Under these conditions, the implantation of heavy ions (Ga) at low energy regime ($30$ keV) is likely to achieve a channeling condition \cite{Schmidt_book_implantation}. 

In order to verify the effect of ionic implantation on the sample structure, Raman measurements were performed. The experiments were carried out with the light incising perpendicular to the sample c-axis, with the electric field polarized along the graphene planes. Results for sample L2 at three implanted doses are shown in the lower pannel of Fig. \ref{fig_impl_diagram}. Such doses correspond to the points marked in the resistance-implantation diagram for all samples, which is shown in the upper pannel of the same figure. The G-peak and 2D peaks on the Raman spectra are fingerprints of graphite, whereas the one labeled D is a signature of structural disorder \cite{Reich_london2004_raman_HOPG}. The curve \# 1 shows the presence of the D-peak due to the disordered carbon layer surrounding the sample (presumably formed during the milling process). For subsequent doses (curves \# 2 and \# 3) the D-peak intensity gets higher as it overlaps with the G-peak. At this stage  ($I_G/I_D \approx 1$) the sample surface reaches the amorphization condition, no longer allowing measurements of the underlying graphite structure.
%(  The ``X'' corresponds to the highest implantation dose achieved before the devices were rendered not measurable)

The Raman spectra can be compared with those previously obtained by B. S. Elman and M. Dresselhaus on HOPG implanted with different ions, albeit parallel to c-axis \cite{Elman_PRB1981_raman_ion_HOPG, Elman_PRB_dmg_ion_HOPG}. For example, ref. \cite{Elman_PRB_dmg_ion_HOPG} shows that the implantation of $100$ keV He ions with a fluence of $1 \times 10^{14} \text{ ions/cm}^2$ is sufficient to produce amorphization of the sample surface. At the same time, lesser energetic ions are shown to produce more damage to HOPG  surface than their higher energetic counterparts \cite{Elman_PRB1981_raman_ion_HOPG}. 

Considering these, it would be expected that the lower energy $\text{Ga}^+$ ions used here ($30$ keV) would not produce an early surface amorphization at fluences of $10^{14} \text{ cm}^{-2}$. This would happen due to the ions' larger energy transfer for collision, smaller scatter angle, smaller range and larger defect density per volume when compared to lighter ions \cite{book_nastasi}. These considerations suggest a channeling scenario for our graphite sample. As the implanted doses gets higher, however, the channels are progressively destroyed as dechanneling and random scattering events become important \cite{GASPAROTTO_Nucl_instr_channeling_N_Si, Venkatesan_JAP1984_recryst_HOPG_surface}. 

Hence, the ever increasing amount of damage can be conciliated with the sample resistance reduction by considering that our devices undergo a progressive amorphization which competes with the doping caused by Ga atoms. The doping can be understood as a consequence of channeled Ga ions acting as an interstitial linking between graphene planes in graphite, which donate electrons \cite{Meng_carbon2013_p_dop_HOPG}. Since the resistivity of graphite depends on the amount of charge carriers \cite{book_kelly}, this contribution reduces the sample resistance according to the inverse of the implanted dose. 

As the progressive damage increases with the amount of implanted ions, however, ionic channeling becomes suppressed. This results in a positive contribution to the sample resistance due to the scattering of upcoming ions, creating cascade process \cite{book_kharisov} which increases the sample amorphization rate. Its contribution to the resistivity is non-trivial, but positive. Assuming that only implanted atoms act as scatters for the upcoming ions, the dependency is linear. Considering these two competing effects, we propose the following phenomenological expression to describe the change of sample resistance with the amount of implanted ions:
\begin{equation}
\text{R}(\text{d})\approx\frac{\text{R}_0}{1+\gamma\frac{\text{d}}{n_0}}+\beta \times \text{d}.
\label{eq_impl}
\end{equation}                           
In it, d is the implanted dose, $\text{R}_0$ is the resistance of the pristine device, $n_0$ the sample's native charge carrier concentration, $\beta$ is a constant representing the resistance gain due to the progressive amorphization and $\gamma$ is an dimensionless constant related the efficiency of the process. The first term in eq. \ref{eq_impl} describes the resistance reduction due to charge doping caused by ions, as the second term describes the disorder-induced (amorphization-related) resistance increase.
%At low doses, the channeling regime dominates. In this stage, the ions penetrate deep into the interplanar region, losing energy by elastic collisions until coming to rest. However, 

This tentative phenomenological model adjusts well to the experimental data with the use of two free parameters $\gamma$ and $\beta$ (except for sample L3, which did not have its $\text{R}_0$ (T = 5 K) measured -- see Fig. \ref{fig_RxT}). The values of $\beta$ and $\gamma$ are within the same order of magnitude for all samples, suggesting that all devices undergo the same process. While the physical meaning of $\beta$ is not evident, the parameter $\gamma$ is directly linked to the efficiency of the implantation. It corresponds to the number of extra carriers added per implanted ion. Assuming ribbons with $n_0 \approx 10^{11}$ $cm^{-2}$ (as in bulk samples) gives an upper limit for $\gamma$ ranging between $3.5 \times 10^{-3}$ and $5.7 \times 10^{-3}$. This corresponds to an efficiency of about $0.5$\% for implanted doses in the range of $10^{14}\text{ cm}^{-2}$. Such small values can be attributed to two factors: the relatively high fluence utilized and the lack of post-implantation thermal annealing. The reduction of the doping efficiency at high fluences is generally attributed to the creation of charge trap defects, which are always induced and hinder transport in the samples \cite{book_kharisov}. This effect is usually remedied by a post-implantation thermal annealing, which was not done here. For example, it is shown in refs. \cite{book_kharisov, GASPAROTTO_Nucl_instr_channeling1995_irr_MgIn2O4, Wagner_PSSA1978_annealing_impl} that thermal annealing in different materials can improve the effective ionic doping above one order of magnitude by prompting recrystallization and solidary chemical bondings between the implanted element and the parent compound. 

\section{Conclusions}

In conclusion, in this work, we have shown ionic doping parallel to planes in HOPG as a viable way to modulate the sample resistivity. Our results can be interpreted as the occurrence of ionic channeling in our samples in competition to amorphization induced by higher fluence ionic implantation. Our results suggest that ionic implantation perpendicular to the c-axis in HOPG modulates the sample charge carrier density while introducing much less damage when compared to implantation parallel to the sample c-axis. Our results point new routes towards the modulation of charge carriers in multigraphene devices and the functionalization of graphite nano objects. 

\section*{Acknowledgments}
We would like to thank Prof. Dr. Walter Escoffier for carefully reading the manuscript, to Prof. P. Esquinazi for insightful comments and to prof. Y. Kopelevich for constructive discussiona and supplying the GW sample. This work was carried out with the financial support of CNpQ (Conselho Nacional de Desenvolvimento Cientifico e Tecnologico - Brazil) and by the Brazilian National Council for the Improvement of Higher Education (CAPES).  B. C. C. acknowledges financial support from the Polish National Science Center under project number UMO-2014/15/B/ST3/03889. We also acknowledge Erasmus Mundus Webb for financial support.

%\bibliography{references_article}
\end{document}